\documentclass{article}
\usepackage{spconf,amsmath,graphicx}

\usepackage{tabu, xcolor, cite}
\usepackage{multirow}
\usepackage{hyperref}

\title{Speaker reinforcement using target source extraction \\for robust automatic speech recognition}

\name{C\u{a}t\u{a}lin Zoril\u{a} and Rama Doddipatla
\thanks{\copyright 2022 IEEE. Accepted for ICASSP 2022.}}
\address{Toshiba Cambridge Research Laboratory, Cambridge, United Kingdom\\
\\\texttt{firstName.lastName@crl.toshiba.co.uk}}

\begin{document}
\ninept
\maketitle
\begin{abstract}

Improving the accuracy of single-channel automatic speech recognition (ASR) in noisy conditions is challenging.
Strong speech enhancement front-ends are available, however, they typically require that the ASR model is retrained to cope with the processing artifacts.
In this paper we explore a speaker reinforcement strategy for improving recognition performance without retraining the acoustic model (AM). This is achieved by remixing the enhanced signal with the unprocessed input to alleviate the processing artifacts.
We evaluate the proposed approach using a DNN speaker extraction based speech denoiser trained with a perceptually motivated loss function.
Results show that (without AM retraining) our method yields about 23\% and 25\% relative accuracy gains compared with the unprocessed for the monoaural simulated and real CHiME-4 evaluation sets, respectively, and outperforms a state-of-the-art reference method.

\end{abstract}

\section{Introduction}
\label{sec:intro}

Recently, the deep neural networks (DNNs) have greatly improved the accuracy of automatic speech recognizers.
The current ASR systems have already reached human performance in clean conditions, however, they are still worse than normal-hearing listeners in noisy conditions~\cite{Trinh_ieeeTASLP21}.

To improve ASR robustness in noise, there are at least two distinct strategies in the literature.
One strategy relies on large amounts of data to train multi-condition models, and the other strategy is on performing data cleaning using signal enhancement.
Although the first approach is simple, it is costly both in terms of computational resources required to train such models and in terms of collecting annotated data. Furthermore, the accuracy of multi-condition systems drops in very challenging environments, such as those with competing speakers~\cite{Barker2018CHiME5}. 
Concerning the data cleaning track, the results reported so far are mixed and they indicate that accuracy gains can be achieved in 
certain conditions (e.g., strong interfering speech~\cite{sato_is2021separate}), but there are shortcomings in others.
Frequently, distortions introduced by speech enhancement limit the applicability of these methods as standalone front-end for ASR, and the acoustic model has to be retrained with matched distorted data to achieve the best recognition performance~\cite{Zorila_asru19_chime5}. Alternatively, jointly trained enhancement and recognition systems have been proposed to alleviate the distortions~\cite{Settle_icassp18,Chang_icassp20}. However, in real applications with very dynamic acoustic conditions, the latter approach may not work well.

In this paper we show that ASR accuracy in noisy conditions can be boosted by using a {\it speaker reinforcement strategy} without acoustic model retraining on distorted data.
Instead of focusing on fully suppressing the background using state-of-the-art enhancement algorithms, we conjecture that by remixing the enhanced signal with the unprocessed input alleviates the processing artifacts, leading to significant recognition accuracy gains without model retraining.
A similar idea has been recently proposed for raw waveform speech denoising in~\cite{defossez20_interspeech}, where a dry/wet knob was demonstrated to improve the overall perceived quality of processed samples, but in that work the ASR experiments did not investigate ASR robustness in mismatched noisy conditions.
Here, we evaluate whether speaker reinforcement is effective for monoaural speech denoising for ASR on both simulated and real noisy data.

A more powerful speech enhancement method is expected to yield better performance, therefore we have chosen a speaker extraction (SPX) system to perform denoising in this work. Instead of recovering all sources from a mixture (i.e., speech separation), the aim of SPX is to recover only a target speaker from the mix~\cite{Zmolikova_asru2017, Xu_asru_2019, giri21_interspeech, zmolikova21_interspeech}, which circumvent the requirement to know in advance the total number of sources. Kinoshita et al.~\cite{Kinoshita_icassp20a} have recently proposed a denoiser based on convolutional time-domain audio separation network (Conv-TasNet,~\cite{Luo_ieee_aslp2019}) to boost single-channel ASR performance in noisy conditions, achieving very promising results on CHiME-4 data. Inspired by those results and also by the success of the time-domain SpeakerBeam SPX algorithm~\cite{Delcroix_icassp_2020, Zorila_slt2021}, which also follows Conv-TasNet's architecture, we selected SpeakerBeam SPX to perform our denoising experiments. We also investigate whether adding knowledge from the perceptual studies during network training strengthens the initial model, as other prior speech enhancement studies have shown~\cite{Kolbaek_icassp2018, saddler21_interspeech, Wang_ieeespl21}.

In summary, our contributions in this paper are the following. We: (i) investigate the ASR performance of time-domain SpeakerBeam for speech denoising of real and simulated data in both matched and mismatched conditions, (ii) propose a new loss function based on an objective intelligibility metric for training time-domain SpeakerBeam denoiser, (ii) suggest speaker reinforcement strategy to improve robustness of ASR models in noisy environments.  
\begin{figure*}[ht]
	\centering
	\includegraphics[width=0.95\textwidth]{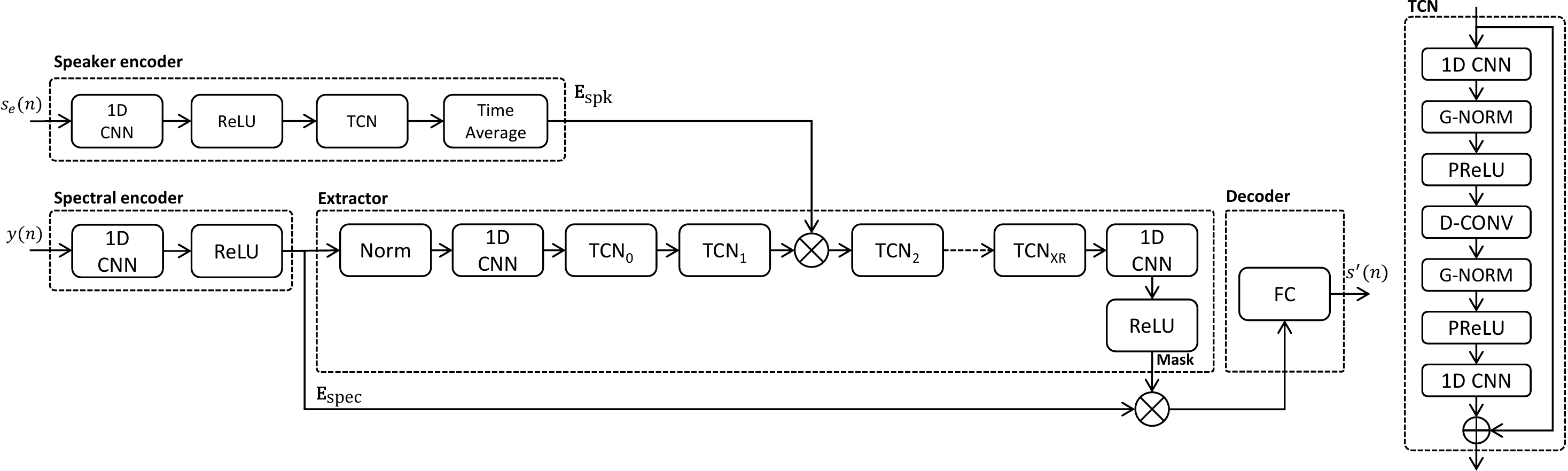}
\caption{Block diagram of single-channel time domain SpeakerBeam~\cite{Delcroix_icassp_2020}.}
\label{fig:spkbeam}
\end{figure*}
The rest of the paper is structured as follows. Section~\ref{sec:spkbeam} briefly presents the time domain SpeakerBeam algorithm for monoaural speaker extraction, Section~\ref{sec:proposed} describes the proposed modified training loss function and the speaker reinforcement strategy, Section~\ref{sec:eval} introduces the evaluation methodology, and Section~\ref{sec:results} shows the results.
The paper is concluded in Section~\ref{sec:concl}.

\section{Single-channel time domain SpeakerBeam}
\label{sec:spkbeam}

Delcroix et al.~\cite{Delcroix_icassp_2020} have recently proposed a mask-based time domain SPX system whose block diagram is depicted in Fig.~\ref{fig:spkbeam}. Single-channel SpeakerBeam is made of four components: a spectral encoder, a speaker encoder for the target, an extractor network and a decoder. Except for the target speaker encoder, SpeakerBeam mainly follows Conv-TasNet's architecture~\cite{Luo_ieee_aslp2019} and it consists of two inputs, one for the mixture and one for the enrolment signal. The later input is used to generate embeddings for the target, which is subsequently employed to bias the extraction.
A brief description of each component is given below.

\subsection{Spectral encoder}

The aim of the spectral encoder is to transform the speech waveform into a higher dimensional space ($\textbf{E}_\mathrm{spec}$) using one CNN layer followed by a rectified linear unit (ReLU) for activation. The CNN has $\mathrm{N}$ kernels of size $\mathrm{L}$ and stride $\mathrm{L/2}$.

\subsection{Extraction network}

The extractor is designed to compute masks for the target speaker by exploiting the sparsity of speech in the spectral domain and the long-term dependencies of its temporal features.
This is achieved by cascading several blocks of temporal convolution networks (TCNs) with increasing kernel dilation factors which perform multi-resolution analysis of input mixture.

As shown in Fig.~\ref{fig:spkbeam}, a TCN block is formed of three CNN layers, parametric ReLU (PReLU) activations and mean and variance normalization across both time and channel dimensions scaled by trainable bias and gain parameters (G-NORM).
The depthwise convolution (D-CONV, $\mathrm{H}$ kernels of size $\mathrm{P}$) operates independently on the input channels, and the dilation factors across consecutive TCN blocks is $2^\mathrm{mod(i,XR)}$, where $\mathrm{i}$ is the block's index, $\mathrm{XR}$ is the total number of blocks and $\mathrm{mod}$ is the modulo operation. The endpoint 1D CNNs ($\mathrm{B}$ kernels) are employed to adjust the channel dimension.
A multiplicative adaptation layer is used to combine the target speaker embedding $\textbf{E}_\mathrm{spk}$ with the output of the second TCN.

The spectral representation $\textbf{E}_\mathrm{spec}$ is firstly channel-wise normalized, then processed by a bottleneck 1$\times$1 CNN layer ($\mathrm{B}$ kernels) before being fed to the first TCN.
Another 1$\times$1 CNN layer with $\mathrm{N}$ output channels is used after the last TCN to adjust the mask dimension to that of spectral encoder's output, thus facilitating their pointwise multiplication.

\subsection{Speaker encoder}

Speaker encoder is made of one CNN layer ($\mathrm{N}$ kernels of size $\mathrm{L}$ and stride $\mathrm{L/2}$), followed by ReLU activation, a TCN configured as described in the previous section (dilation 1), and a time averaging operator.

\subsection{Decoder}

The decoder reconstructs the estimated target frames $s'(n)$ using the masked spectral representation and one fully-connected layer ($\mathrm{N}$ input and $\mathrm{L}$ output dimensions). Overlap-and-add is applied to reconstruct the whole waveform.

\subsection{Training criterion}

The training objective for SpeakerBeam is to maximize the scale-invariant signal-to-distortion ratio (SISDR), which is defined as:
\begin{equation}
\label{eq:sisdr}
\mathrm{SI \text - SDR}=10 \log_{10}{\frac{\left\| \frac{\left\langle s',s \right\rangle}{\left\langle s,s \right\rangle} s \right\| ^2}{\left\| \frac{\left\langle s',s \right\rangle}{\left\langle s,s \right\rangle} s - s' \right\|^2}}
\end{equation}
where $s'$ and $s$ denote the estimated and the oracle target speaker signals, respectively.

\section{Proposed method}
\label{sec:proposed}

Preliminary experiments have shown that the distortions produced by the SPX processing harm the ASR accuracy, especially in mismatched scenarios, therefore we have explored two strategies to alleviate them, as described next.

Firstly, a novel training criterion is proposed that combines the standard SISDR loss with a perceptually motivated term based on the short-term objective intelligibility (STOI) measure~\cite{Taal_icassp2010}:
\begin{equation}
\mathrm{L_ \text {new}} = \mathrm{L_ \text {SISDR}}(s,s') + \mathrm{L_ \text {STOI}}(s,s'),
\end{equation}
where $\mathrm{L_ \text {SISDR}}$ is the standard SpeakerBeam loss defined in Eq.~(\ref{eq:sisdr}), and the second term is the STOI-based loss.
STOI is a widely used metric to objectively assess the intelligibility of noisy speech processed by time-frequency weighting, and it was shown to produce a high correlation with human perception, thus it could also help reducing signal distortions for ASR. A DNN based speech enhancement system that maximizes an approximation of STOI has already been proposed in the literature~\cite{Kolbaek_icassp2018}, which, however, achieved modest gains compared with classical mean square error DNN systems. Instead of relying exclusively on STOI to train the denoiser, here we combine it with the SISDR loss. To the best of our knowledge, this is a novel approach in the context of SPX enhancement and for robust ASR applications.
\begin{figure}[ht]
	\centering
	\includegraphics[width=0.8\columnwidth]{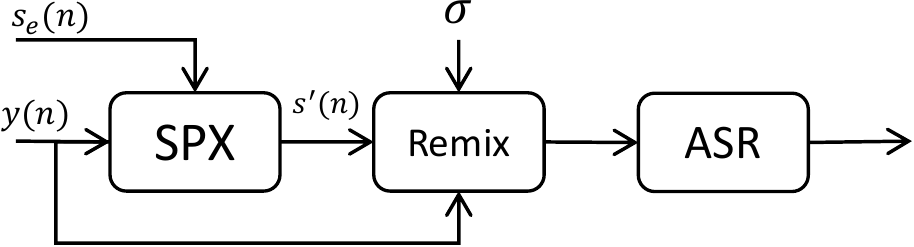}
\caption{Proposed target speaker reinforcement method.}
\label{fig:proposed}
\end{figure}

Secondly, we propose remixing the enhanced signal with a fraction of the input mixture to mask the processing artifacts that could harm ASR. We denote this strategy as {\it target speaker reinforcement}. The remixing is controlled by a scaler $\sigma$~(Fig.~\ref{fig:proposed}):
\begin{equation}
\sigma = 10 \log_{10} \frac{\left\|s' \right\|^2} {\left\| \alpha y \right\|^2},
\end{equation}
and the output is computed as $z(n) = s'(n) + \alpha y(n)$. The role of $\sigma$ on the ASR accuracy of a pre-trained acoustic model has been investigated in the evaluation section.

\section{Evaluation}
\label{sec:eval}

Experiments have primarily focused on the CHiME-4 data, which contains both simulated and real noisy speech recordings~\cite{Barker_csl2017}.
The CHiME-4 corpus was designed to capture multi-channel speech using a mobile tablet computing device in noisy everyday environments such as cafeteria, on the bus, street or pedestrian areas. The SPX denoiser is trained on clean Wall Street Journal (WSJ) speech artificially mixed with CHiME-4 noise.
We report results on the single-channel real and simulated evaluation set ($\mathrm{et05}$) of CHiME-4 (matched conditions with respect to SPX's train set).
Additionally, we also report results in mismatched test conditions for the VoiceBank-DEMAND (VBD) and WHAM! sets~\cite{Botinhao_2017, Wichern_is2019}. The max version of WHAM! test ($\mathrm{tt}$) set was used, and all experiments were performed with 16~kHz resolution data.

Proposed method is compared with two reference methods. 
One is MetricGAN+~\cite{fu_is21_metricGanPLus}, which is a state-of-the-art single channel speech denoiser, and the other one is a TasNet based single-channel speech denoiser (Denoising-TasNet,~\cite{Kinoshita_icassp20a}). The SpeechBrain\footnote{https://speechbrain.github.io} implementation of MetricGAN+ was employed for our evaluation.

The performance is mainly assessed in terms of word error rate (WER), however, for some preliminary experiments,
the signal-to-distortion (SDR) and STOI values are reported as well. The SDR scores are computed using the BSSeval toolkit~\cite{Vincent_ieee_aslp2006}, and the STOI training loss is computed using a freely available PyTorch implementation\footnote{https://github.com/mpariente/pytorch\_stoi}. More details about the configuration of the denoising network and ASR systems are presented next.

\subsection{Speaker extraction based denoiser}

Single-channel time-domain SpeakerBeam proposed by Delcroix et al.~\cite{Delcroix_icassp_2020} is used to perform speech denoising, with
the same configuration as described in~\cite{Zorila_slt2021}, but without the spatial encoder.
The spectral encoder consists of 1-D CNN with $\mathrm{N}$=256 kernels of size $\mathrm{L}$=20 and a frame rate of 10 samples.
$\mathrm{X}$=8 stacked TCN blocks repeated $\mathrm{R}$=4 times are employed for the extraction network. Each TCN block consists of 1$\times$1 CNNs and 1$\times$3 depthwise convolutions with $\mathrm{B}$=256 and $\mathrm{H}$=512 kernels, respectively.
The fully connected layer in the decoder has an input dimension of 256 and an output dimension of 20.

The clean WSJ training list from WSJ0-2mix~\cite{Hershey_icassp2016} artificially added with CHiME-4 noise was used to train the SPX system. About 39~hours of  data have been generated, the signal-to-noise (SNR) ratio of the mixtures being uniformly sampled from a range of 0~dB to 5~dB, and the audio length randomly varied from 1~s to 6~s. The target enrolment sentences from~\cite{Xu_asru_2019} have been used for training, making sure that the recordings for enrolment and mixture signals were different. The enrolment samples for the simulated CHiME-4, VBD and WHAM! test sets were chosen from the available clean waveforms, while the close-talk microphone recordings were used for enrolment of the real CHiME-4 evaluation set.

Training was performed using the Adam optimizer~\cite{Kingma_iclr2015} with the initial learning rate of 0.001, a chunk length of $\mathrm{4}$~seconds, and a minibatch size of $\mathrm{8}$. The learning rate halved if no improvement on the cross-validation set was yielded for three consecutive epochs. All competing models were decoded at epoch 20 to avoid overfitting on the training data.

\subsection{Automatic speech recognition}

Two acoustic models are included in the evaluation. The first model is trained on clean WSJ-SI284 data (WSJ-CLN) and has a 12-layer TDNNF topology~\cite{Povey_is2018_tdnnf}, while the second model is trained on the standard noisy set from CHiME-4 (C4-ORG) and has a 14-layer TDNNF structure.
The later system employs all six channels from the real and simulated training sets of CHiME-4.
Both models use 40-dimension MFCCs and 100-dimension i-vectors as acoustic features, and they were trained in KALDI using the lattice-free MMI criterion~\cite{Povey_is2016}. Both a standard tri-gram and a more powerful RNN language model (LM) are used for decoding.
After 3-fold speed perturbation, WSJ-CLN and C4-ORG have about $\mathrm{246}$ hours and $\mathrm{327}$ hours of training data, respectively.

\section{Results \& Discussion}
\label{sec:results}

This section presents the results of our investigation into the effectiveness of proposed target speaker reinforcement approach for improving ASR robustness in matched and mismatched noise conditions.

\begin{table*}[h]
\centering
\small
\setlength{\tabcolsep}{4pt}
\caption{Performance of proposed method in mismatched noisy conditions (Denoising-SPX was trained on simulated noisy CHiME-4 data). WER (\%) are with WSJ-CLN AM and 3-G LM.}
\label{tab:wsj_clean}
\begin{tabu}{lccccc|ccc}
\hline
\multirow{2}{*}{Enhancement} & \multirow{2}{*}{Train Loss} & \multirow{2}{*}{\shortstack{$\sigma$ \\(dB)}} & \multicolumn{3}{c|}{VBD Test} & \multicolumn{3}{c}{WHAM! tt} \\ \cline{4-9}
& & & \multicolumn{1}{c}{WER(\%)} & \multicolumn{1}{c}{SDR(dB)} & \multicolumn{1}{c|}{STOI} & \multicolumn{1}{c}{WER(\%)} & \multicolumn{1}{c}{SDR(dB)} & \multicolumn{1}{c}{STOI} \\ \hline \hline
\multirow{1}{*}{Unprocessed} 			& -	& - & 38.6 & 8.5 & 0.92  & 75.7 & -2.8 & 0.76 \\ 
\multirow{1}{*}{MetricGAN+~\cite{fu_is21_metricGanPLus}} 			& -	& - & 35.4 & 15.0 & 0.93  & 77.5 & 1.8 & 0.72 \\ 
\multirow{1}{*}{Denoising-SPX (vanilla)} & SISDR & $\infty $ & 33.7 & 16.1 & 0.92  & 25.4 & 11.7 & 0.92 \\
\multirow{1}{*}{Denoising-SPX (proposed)}& SISDR+STOI& $\infty$ & {\bf 33.2} & 15.8 & 0.93  & {\bf 24.6} & 11.6 & 0.92 \\ \hline
\rowfont{\color{gray}}
\multirow{1}{*}{Clean}				& -	& - & 25.1 & $\infty$ & 1.00  & 9.0 & $\infty$ & 1.00  \\ \hline
\end{tabu}
\end{table*}

Firstly, the SPX denoiser is evaluated in mismatched conditions using VBD and WHAM! simulated noisy test data (Table~\ref{tab:wsj_clean}).
The WER results in Table~\ref{tab:wsj_clean} are with the WSJ-CLN AM and a tri-gram (3-G) LM.
Although the Denoising-SPX is trained on simulated noisy CHiME-4 mixtures and, therefore, is mismatched with either test sets, it yields about 14\% and 67\% relative WER reduction on the VBD and WHAM! test sets compared with the unprocessed case. Notably, Denoising-SPX decisively outperforms the reference system MetricGAN+ (trained on VBD data) in both cases. MetricGAN+ achieved a worse WER than the unprocessed for the WHAM! set, indicating that the current pre-trained system cannot cope with unseen noise conditions.
WER results in Table~\ref{tab:wsj_clean} show that the composite SISDR and STOI training loss works better than the standard SISDR loss, although the SDR and STOI values for the vanilla and proposed systems are almost identical. We believe that the additional STOI term is able to help restore some of the temporal modulations of speech distorted during enhancement.
{Our preliminary evaluation showed that training a STOI only denoising system leads to worse performance than by using the pure SISDR loss. A more complete analysis on 
weighting the STOI and SISDR loss is planned for our future work.}

Next set of experiments are performed using the noisy CHiME-4 acoustic model (C4-ORG) and they are assessing the importance of the remixing ratio $\sigma$ for ASR robustness. Results in Table~\ref{tab:chime4_remix} show that impressive WER reductions can be achieved by decreasing the value of $\sigma$ from $\infty$ (no input mixture is added on top of the enhanced signal) to $\mathrm{0}$~dB for both the simulated and real CHiME-4 test sets. {Reducing $\sigma$ further yields a reversing (increasing) trend for the ASR accuracy, as shown in Table~\ref{tab:chime4_remix}.}
Only by decreasing the remixing ratio, the proposed Denoising-SPX achieves about 28\% and {36}\% relative WER reduction for the simulated and real evaluation set, respectively.
These results are remarkable since neither the acoustic nor the SPX models were retrained to achieve these gains. The poor performance of Denoising-SPX for $\sigma=\infty$ compared with the unprocessed case can be attributed to the fact that the system was trained from anechoic simulated CHiME-4 noisy data, while the test sets also contain a small amount of reverberation. Another source of accuracy degradations could be the inherent distortions introduced by SPX, especially with the real data.

\begin{table}[h]
\centering
\small
\setlength{\tabcolsep}{6pt}
\caption{WER accuracy of proposed speaker reinforcement approach on CHiME-4 for various remixing ratios, $\sigma$. All results are with noisy CHiME-4 AM (C4-ORG, 3-G LM).}
\label{tab:chime4_remix}
\begin{tabu}{lccc}
\hline
\multirow{2}{*}{Enhancement} & \multirow{2}{*}{\shortstack{$\sigma$ \\(dB)}} & \multicolumn{2}{c}{WER(\%)} \\ \cline{3-4}
& & \multicolumn{1}{c}{et05\_simu} & \multicolumn{1}{c}{et05\_real} \\ \hline \hline
\multirow{1}{*}{Unprocessed} 				& - & 16.4 & 15.5 \\ \hline
\multirow{4}{*}{\shortstack{Denoising-SPX \\(vanilla)}} & $\infty$ & 18.4 & 19.4 \\
																												& 20 & 16.5 & 17.7 \\
																												& 10 & 14.9 & 15.5 \\
																												& 0  & 13.5 & 13.0 \\ 
																												& -10  & 13.7 & 12.2 \\
																												& -20  & 14.9 & 13.2 \\ \hline
\multirow{4}{*}{\shortstack{Denoising-SPX \\(proposed)}}& $\infty$ & 18.4 & 18.9 \\
																												& 20 & 16.4 & 17.2 \\
																												& 10 & 14.4 & 15.1 \\
																												& 0 & {\bf 13.2} & 12.7 \\
																												& -10 & 13.7 & {\bf 12.0} \\
																												& -20 & 14.8 & 13.1 \\ \hline
\rowfont{\color{gray}}
\multirow{1}{*}{Clean}					& - & 2.5 & 4.9  \\ \hline
\end{tabu}
\end{table}

\begin{table}[t]
\centering
\small
\setlength{\tabcolsep}{5pt}
\caption{WER accuracy of proposed method on CHiME-4 using C4-ORG AM and RNN LM.}
\label{tab:chime4_all}
\begin{tabu}{lccccc}
\hline
\multirow{3}{*}{Enhancement} &  \multirow{3}{*}{\shortstack{$\sigma$ \\(dB)}} & \multicolumn{4}{c}{WER(\%)} \\ \cline{3-6}
 & & \multicolumn{2}{c}{et05\_simu} & \multicolumn{2}{c}{et05\_real} \\ \cline{3-6}
 & & \multicolumn{1}{c}{3-G} & \multicolumn{1}{c}{RNN} & \multicolumn{1}{c}{3-G} & \multicolumn{1}{c}{RNN} \\ \hline \hline
{Unprocessed} 	
																& - & 16.4 & 13.6 & 15.5 & 12.3 \\ \hline
{Denoising-TasNet~\cite{Kinoshita_icassp20a}}
																& - & - & 11.9 & - & 9.8 \\ \hline
\multirow{2}{*}{\shortstack{Denoising-SPX (proposed)}}
																& $\infty$ & 18.4 & 14.4 & 18.9 & 14.9 \\
																& 0 & 13.2 & {\bf 10.4} & 12.7 & 9.6 \\ 
																& -10 & 13.7 & 10.8 & 12.0 & {\bf 9.2} \\ \hline
{Denoising-SPX (proposed)*}			& 0 & 12.9 & {\bf 9.9} & 11.7 & {\bf 8.6} \\ \hline
\rowfont{\color{gray}}
{Clean}					& - & 2.5 & 1.5 & 4.9 & 3.2 \\ \hline
\end{tabu}
\end{table}

Table~\ref{tab:chime4_all} compares the recognition accuracy of Denoising-SPX with that of Denoising-TasNet~\cite{Kinoshita_icassp20a} on the single-channel CHiME-4 task. To ensure a fair comparison with the results reported in~\cite{Kinoshita_icassp20a}, the standard noisy C4-ORG acoustic model is used to perform ASR, and the 3-G transcriptions are rescored with an RNN-based language model. Notably, the proposed Denoising-SPX with speaker reinforcement has outperformed Denoising-TasNet in both simulated and real evaluation sets by 13\% and {6}\% relative WER, respectively.
This is a remarkable outcome since Denoising-SPX is trained on anechoic simulated CHiME-4 noisy data, while the Denoising-TasNet had to be trained on simulated reverberated and noisy recordings because the performance was not sufficient without reverberation~\cite{Kinoshita_icassp20a}.
Compared with the Unprocessed for RNN-LM, Denoising-SPX with speaker reinforcement yielded about 23\% and {25}\% relative WER reduction for simulated and real sets, respectively.
The second-to-last line in Table~\ref{tab:chime4_all} shows the performance of the proposed system trained using also artificial reverberation that matches CHiME-4 conditions. Additional accuracy improvements are observed for both the simulated and real test sets.

As future work, we plan to extend the evaluation of proposed method in reverberant, noisy and multi-talker conditions.

\section{Conclusions}
\label{sec:concl}

In this paper we have presented a target speaker reinforcement algorithm for improving the ASR accuracy in noisy conditions without acoustic model retraining.
Using a denoiser based on a DNN speaker extraction, we show that remixing the noisy input with the enhanced signal achieves about 23\% and {25}\% WER reduction compared with the unprocessed case for the single-channel CHiME-4 simulated and real evaluation sets, respectively.
Furthermore, the experiments suggest that adding a perceptually motivated loss on top of a time domain reconstruction loss during training of speaker extraction systems, helps achieve a modest but consistent ASR accuracy gain.

% -------------------------------------------------------------------------
\bibliographystyle{myIEEEbib}
\bibliography{IEEEabrv,mybib}

\end{document}